\documentclass[preprint,authoryear,12pt]{elsarticle}
\usepackage{subfigure}
\usepackage{amssymb}
\journal{Advances in Space Research}

\begin{document}

\begin{frontmatter}

\title{Long-term temporal variations in the areas of sunspot groups}

\author{J.\ Javaraiah}

\address{Indian Institute of Astrophysics, Bengaluru - 560034, India \\
Tel: +91 80 25530672,  Fax: +91 80 25534043}

\ead{jj@iiap.res.in}

\begin{abstract}
{\small Recently, Javaraiah (2012a) analyzed the combined Greenwich 
and Solar Optical Observing 
Network (SOON)
 sunspot group data during the period 1874\,--\,2011 
 and studied variations in the annual numbers (counts) of the small
 (maximum area $A_{\rm M} < 100$ millionth of solar hemisphere, msh), 
large ($100 \le A_{\rm M} < 300$ msh),  
and big ($A_{\rm M} \ge 300$ msh) sunspot groups. Here  that 
analysis is extended and studied variations in the mean maximum sizes 
(the mean values of  maximum areas) of 
 the aforementioned three classes of sunspot groups 
and also  their combination. 
 It is  found that there is no significant  correlation between the 
mean maximum size of 
any class of sunspot groups and the International 
Sunspot Number ($R_{\rm Z}$), probably due to  
in a given time interval  small sunspot groups/sunspots outnumber the
 large ones.
 A pattern of  an approximate 9-year period
 cycle is seen in   the variations of the mean maximum sizes of the large 
and the big sunspot  groups during  a  solar cycle.
On long-time scales it is found that  there exists
 a strong  130 or more years cycle in the
 variation of the mean 
size of the small sunspot groups, whereas there is a hint on the existence of 
$\approx$ 44-year cycles
in the variations of the mean maximum sizes of the large and the big sunspot
 groups.
 During the decline phase of cycle~23, there was a  scarcity in 
the sunspot groups whose $A_{\rm M} \le$37 msh, which may be related to the 
  slow growth    of
sunspot groups during this period. 
 During the minimum between cycles 23 and 24 
  may be due to the presence of a
  number of small 
sunspot groups  
whose $A_{\rm M} > 37$ 
was larger than that of whose  $A_{\rm M} \le$37,  
the relatively large size 
coronal holes were present  at low-latitudes  and the total solar irradiance
 was very low.}
\end{abstract}
\begin{keyword}
solar magnetic field \sep solar activity \sep solar cycle
\end{keyword}
\end{frontmatter}
\parindent=0.5 cm

%\newpage
\section{Introduction}
The study of variations in solar activity is important for understanding 
the  mechanism behind the solar activity,   
and because of solar activity affects
 space weather and it may also have a contribution in Earth's 
climate change~\citep[e.g.,][]{dh10,ay12}.
Although it is well believed that the solar cycle results from generation of 
strong toroidal magnetic fields in the Sun, with a main period of 
about 22-years, by combined effects of 
convection and differential rotation in the Sun~\citep[$e.g.$,][]{dg06}, 
 it is not known exactly what causes solar cycle. 
 Because of the long record of
 observations, sunspot measurements
 constitute a primary source of information to better understand 
the level and nature of solar activity.
It is well-known that the dynamic properties, such as rotation rate and  
meridional motion,  of sunspot groups  depend on 
their sizes.  This may 
 imply the dynamics of the Sun's subsurface layers 
  in which the different size   sunspot groups 
anchor/generated~\citep{how96,jg97,jj99,kmh02,siva03,siva07,siva10,sg04}.
Therefore,  studies on the 
variations 
in the  sizes of sunspot groups are important,  which may provide 
important clues on the underlying  mechanism of  solar activity 
 and 
its variations. In some studies it is found that the 
 sunspot area distribution  does not  depend on the Z\"urich 
or International Sunspot Number ($R_{\rm Z}$)~\citep[$e.g.$,][]{babij11}.
Recently, 
 we have analyzed 
the combined Greenwich 
and  SOON
 sunspot group data during the period 1874\,--\,2011 and studied   
 variations in the annual numbers (counts) of the small, 
large,  
and big  sunspot groups, by classifying the sunspot groups on the basis  of 
their   maximum areas \citep[][hereafter Paper~I]{jj12a}. Although, it is found that
  the number of the sunspot groups in each class  shows 
the 11-year periodic variation of  solar cycle, 
 some noticeable differences are seen in  the  
  variations of the numbers of the sunspot group in different classes.
 In the present paper  we 
extended that analysis and studied long-term variations in the mean maximum
 size of each of 
 the three classes of sunspot groups 
and also that of all the sunspot groups in the combined class of 
the three classes.

In the next section we will describe
  the data and the method of analysis. In Section~3 we will
  present the results
 and in Section~4 we will  present conclusions and a brief discussion.

\section{Data and Analysis} 

Here the data  and the method of analysis are the same as in 
Paper~I. 
 We have used the combined  Greenwich and SOON
sunspot group data during the period  May-1874 to May-2011
(taken from \break {\tt http://solarscience.msfc.nasa.gov/greenwch.shtml}).
 These data
include the observation time (the Greenwich data contain the date with the
fraction of a day, in  the SOON data
 the fraction  is rounded to 0.5 day),
heliographic latitude  and
longitude, central meridian distance (CMD), and
corrected umbra and whole-spot areas (in msh), {\it etc.}, of sunspot
groups for each day of observation.
 The
positions of the sunspot groups  are geometrical  positions of the
centers of the groups.

The Greenwich data (May-1874 to December-1976) have been compiled
 from the majority of the white
light photographs which were secured at the Royal Greenwich Observatory
and at the Royal Observatory, Cape of Good Hope. The gaps in their
observations were filled with photographs from  other observatories,
Cape Town, South Africa; Kodaikanal, India; and Mauritius.
The SOON data (January-1977 to May-2011)  include measurements made   by the
United States Air Force (USAF) based on 
 sunspot drawings obtained by a network of the observatories
 in Boulder,  Hawaii, and so on.
David Hathaway  scrutinized
the Greenwich and SOON data and produced a reliable
continuous data series
from 1874 up to date.
 In the case of SOON data, we increased the corrected whole-spot
 areas by a factor of 1.4.
 David Hathaway found this correction was necessary to
 have a combined homogeneous
 Greenwich and SOON
data. However, the  agreement between these two data sets may be still
 not 100\% ~\citep{hc08}.
The combined  Greenwich and SOON sunspot group data  are the largest
 available, 
 reliable data that include  
 the  positions and the areas of sunspot groups. 

We have used the data on only those sunspot groups whose birth and
 death occurred within a disk passage. 
 That is, we have not used the  sunspot groups whose central meridian distance 
 $|{\rm CMD}| > 75^\circ$ in any day of their respective life times. This reduces 
the foreshortening effect and helps to obtain the maximum area of a  sunspot
 group unambiguously.

  If $A_1$, $A_2$,\dots,$A_n$ denote the areas (corrected for the 
foreshortening effect) of all the sunspots in  a
 sunspot group  observed at times 
$t_1$, $t_2$,\dots,$t_n$  during the life time of the 
sunspot group  $T = t_n - t_1$ (days), then  the  
maximum area  is defined as follows:  
$$A_{\rm M} = {\rm max} (A_1, A_2,\dots,A_n)\ , \eqno(1)$$
\noindent  where $n = 2, 3, \dots$.  
Each appearance of a recurrent group is treated 
as an independent group. 
Thus, $T \le 12$ days.
We have used here only the sunspot groups which had $T \ge 1$ day (SOON 
data do not contain the data on the sunspot groups whose $T < 1$ day).   

On the basis of  $A_{\rm M}$ values   we have classified sunspot groups 
 into three classes  as  
 small sunspot groups (SSGs: $A_{\rm M} < 100$ msh),
large sunspot groups (LSGs: $100 \ge A_{\rm M} < 300$ msh), 
and big sunspot groups (BSGs: $A_{\rm M} \ge 300$ msh).
We determined the mean maximum sizes 
(sum of the maximum areas divided by the number of groups)
 $\bar A_{\rm SSG}$, $\bar A_{\rm LSG}$, and  $\bar A_{\rm BSG}$   
 of SSGs, LSGs,  and  BSGs respectively,  for each year during the
 period 1874\,--\,2011 as follows:
$$\bar A_{\rm SSG} = \frac{1}{\rm NSG} \sum^{\rm NSG}_{i = 1} A_{{\rm M},i}\ ,\ 
 \bar A_{\rm LSG} = \frac{1}{\rm NLG} \sum^{\rm NLG}_{j = 1} A_{{\rm M},j}\ ,\ 
  {\rm and} $$  
$$ \bar A_{\rm BSG} = \frac{1}{\rm NBG} \sum^{\rm NBG}_{k = 1} A_{{\rm M}, k}\ , \eqno(2)$$ 

\noindent where 
 NSG, NLG,  and NBG   are
the numbers (counts)   of  SSGs, LSGs, and BSGs, respectively,
in a given year (${\rm NSG} > {\rm  NLG} > {\rm NBG}$). The mean maximum 
size ($\bar A_{\rm ASG}$) of all sunspot groups (ASGs) 
 of the same year is calculated as follows: 
$$\bar A_{\rm ASG} =\frac{1}{{\rm NSG} + {\rm NLG} + {\rm NBG}} 
({\rm NSG} \times \bar A_{\rm SSG} +  {\rm NLG} \times \bar A_{\rm LSG} +
 {\rm NBG} \times \bar A_{\rm BSG})\ .  \eqno(3) $$
We have also calculated the mean maximum size  of  each of the above 
 classes of sunspot groups over a solar cycle ($\bar A_{\rm WCN}$, the sum of the 
yearly means divided by the number of years of a cycle)
  as follows: 
$$\bar A_{\rm WCN} =\frac{1}{L_{\rm WCN}} \sum^{L_{\rm WCN}}_{i = 1} 
\bar A_{{\rm c}, i}\ ,  \eqno(4) $$
\noindent where WCN and  $L_{\rm WCN}$ represent the Waldmeier cycle number 
and  length (in years) of a cycle, respectively;  c represents  any 
 of the above classes of sunspot groups, and $i$   is 
 a year of the corresponding cycle. The uncertainty (standard error)
 in 
$\bar A_{\rm WCN}$ is also determined.

We determined the variations in  
$\bar A_{\rm SSG}$, $\bar A_{\rm LSG}$,  $\bar A_{\rm BSG}$,
 $\bar A_{\rm ASG}$, and $\bar A_{\rm WCN}$.    
    
\section{Results}

Fig. 1 shows the  variations in 
$\bar A_{\rm SSG}$, $\bar A_{\rm LSG}$,   $\bar A_{\rm BSG}$,
 and $\bar A_{\rm ASG}$ ($i.e.$,     
annual variations in the mean maximum 
sizes
 of    SSGs, LSGs, BSGs, and ASGs, $cf.$, Eqs.~(2) and (3)) 
 during the period 1874\,--\,2011 
(a value equal to zero implies  the absent of sunspot groups in 
the corresponding class).  
 In the same figure we have also shown 
the variation in the annual mean
  $R_{\rm Z}$, whose values are      
 taken from the website, 
{\tt ftp://ftp.ngdc.noaa.gov/STP/SOLAR\_DATA/\break SUNSPOT\_NUMBERS/INTERNATIONAL/yearly/YEAR.PLT}).  As can be seen in this figure the mean 
maximum size of the sunspot groups in each class
  varies considerably, but the patterns of the 
 variations considerably differ with the patterns of the corresponding 11-year 
cycles in $R_{\rm Z}$.  There is no 
significant correlation between the mean maximum size of 
any class of  sunspot groups and $R_{\rm Z}$ (in the case of big groups the
value of the 
 correlation coefficient is largest and equal to 0.315).
The mean maximum size of the sunspot groups 
in  any 
 class does not have a 11-year solar cycle pattern. 
 Figs.~2 and 3 show the comparisons of 
$\bar A_{\rm SSG}$ and $\bar A_{\rm LSG}$ with $R_{\rm Z}$ 
 during different solar cycles, 
using  superposed epochs related to the respective
maximum epochs of the cycles (corresponding figures of   
 $\bar A_{\rm BSG}$ and  $\bar A_{\rm ASG}$ are not shown).
As can be seen in Fig.~3, there is a pattern of  an approximate 
9-year cycle  in 
 $\bar A_{\rm LSG}$. The similar pattern is also found in the case of
   $\bar A_{\rm BSG}$,  but
 in  $\bar A_{\rm SSG}$ it cannot be seen clearly (see Fig.~2). 
  No  clear such  pattern seen 
 in  $\bar A_{\rm ASG}$ also. 
 That is, in both of  these cases  the variations
 during different 
solar cycles are  highly  different. 
The mean variation (represented by the solid curve)  is statistically highly 
insignificant, $i.e.$ the errors bars are too  large.

   Figs.~4(a)\,--\,4(d) show the variations in 
$\bar A_{\rm WCN}$ ($cf.$, Eq.~(4))   of the sunspot groups in each class
 ($i.e.$, cycle-to-cycle 
variations in the mean maximum sizes of SSGs, LSGs, BSGs and ASGs).
 As can be seen in this figure   $\bar A_{\rm WCN}$    
 of SSGs  varies systematically on a time scale
 of about 12 cycles, 
suggesting the existence  of a strong  130 or more years  cycle 
in $\bar A_{\rm WCN}$,     whose minimum 
 was taken place  during  cycles~15\,--\,16.
 The long-term variations in $\bar A_{\rm WCN}$  of LSGs and BSGs 
suggest the existence of $\approx$ 44-year cycles in these parameters, 
but the variations are 
not well defined (the  error  bars are large). 
 A  44-year periodicity was seen in sunspot activity~\citep{jj08}.
 From cycle~12 to 20 the pattern of the 
variation in  $\bar A_{\rm WCN}$  of ASGs 
(see Fig.~4(d))  is 
largely  same as the corresponding variations in  
 $\bar A_{\rm WCN}$  of SSGs, whereas it is drastically different during 
the last three cycles in which it seems the sunspot groups were relatively small 
 (this could be due the  
 areas of sunspot groups of SOON data  may not be increased adequately,
 see Section~3).
The sunspot groups in cycles~12 and 17 were  relatively 
 large, whereas in cycles~14  and 20 (relatively small and long cycles)
 they were relatively small, suggesting  the  existence of a 55\,--\,60 
year cycle in  $\bar A_{\rm WCN}$  of ASGs.

Incidentally,  the values of $\bar A_{\rm WCN}$ of SSGs of cycles~12 and 23 are equal (see Fig. 4(a)). In the case of cycle~23, the high value of
 $\bar A_{\rm WCN}$ of SSGs 
 was largely contributed by the sunspot groups in the decline phase, whereas in 
case of cycle~12
  the high value of  $\bar A_{\rm WCN}$ of SSGs was 
largely contributed by  the sunspot groups in the rise phase (see Fig.~1).
 In fact, in the former case the
  dominance in the mean maximum sizes of SSGs  was  continued from the
 decline phase of cycle~11.
 During this period of cycle~11 the mean maximum sizes of all
 the three classes
of sunspot groups were relatively larger, whereas in the case of the
 decline phase of  cycle~23 only 
the mean maximum sizes of SSGs  were relatively larger. That is, 
during this period of cycle~23 (and some extent also during the rise 
of cycle~12) 
there was a scarcity in 
the sunspot groups whose  areas
 $\le$37 msh. This can be seen in Fig. 5, in which
solar cycle variations in the annual numbers of the small sunspot groups
 whose $A_{\rm M} \le 37$ msh  and $37 < A_{\rm M} \le 100$ msh 
are shown. During 2000\,--\,2006 the values that are represented by the
 red squares (connected by dotted curves) in 5(a) are  lower than the corresponding values in 5(b). 
   Fig.~6 shows the cycle-to-cycle variations in the mean annual
 numbers of  these two classes of the small sunspot groups. 
As can be see  in this figure 
in cycle~23 on the average the annual mean number of sunspot groups whose 
$A_{\rm M} \le 37$ msh is relatively  less than those of whose
  $37 < A_{\rm M} \le 100$ msh.

Cycle~23 was unusually long (about 12.4 years) and followed by 
the unusually deep and prolonged minimum. 
When the Sun reached the minimum activity between cycles~22 and 23 in 1996,
 it seems the  low-latitude coronal holes were almost completely absent, whereas in 2007 and 
2008, in spite of the extremely low sunspot activity, coronal holes at 
low-latitudes were still relatively large~\citep[e.g.][]{dtoma11}.  
The occurrence of BSGs and LSGs stopped in 2004 and 2006, respectively, 
whereas the  occurrence of SSGs continued~\citep[see Fig. 1, also see][]{jj12a}.
Fig.~7 shows the plot of yearly mean sizes of SSGs 
 during the preceding minimum epochs of cycles~12\,--\,23 and around the 
 minimum between cycles~23 and 24 ($i.e.$, in years 2006, 2007, 2008
 and 2009) versus the corresponding years.
 As can be seen in this figure in 2007 the mean size of SSGs
 whose   $37 < A_{\rm M} \le 100$ msh is relatively large, 
 and in  2008 the overall mean size of
 all SSGs are relatively large. In 2007 the mean sizes of both the two 
classes of SSGs are relatively larger than corresponding values of in 2008,
 whereas in the case of the combined class of SSGs the value at 2008 is
relatively large 
 (this is because the number of sunspot groups with  $A_{\rm M} \le 37$  are 
relatively less in 2008).  
 The presence of relatively large coronal holes at low-latitudes
 in 2007 and 2008 may be 
related to the presence of relatively large size  SSGs in these years. 
Since the sunspots block the the Sun's radiative output,  
the existence of the relatively 
large size SSGs in the minimum between 
cycles~23 and 24~\citep[the evolution rate of SSGs is also 
relatively low,][]{jj12b}    
my be also responsible for the existence of very low total solar 
irradiance  during this period~\citep{fro09}.

\section{Conclusions and Discussion}
From the above analyses of  a large set of sunspot group data 
 the following  conclusions can be drawn: 
\begin{enumerate}
\item  There is no 
significant correlation between the yearly mean maximum size of 
 the sunspot groups in any class and $R_{\rm Z}$.
\item A pattern of $\approx$ 9-year cycle is seen  in the 
 variations of   the mean maximum
sizes of LSGs and BSGs with phases of solar cycles.
\item There exists a strong  130 or more years cycle in the variation of the
 mean maximum  
size of SSGs, whereas there is a hint on the existence of 
$\approx$ 44-year cycles
in the variations of the mean maximum sizes of LSGs and BSGs. 
\item  During the decline phase of cycle~23, there was a   scarcity in 
the sunspot groups whose areas $\le$37 msh. 
\item During the minimum between cycles 23 and 24 
  may be due to the presence of a  number of small sunspot groups  
whose $A_{\rm M} > 37$ 
was larger than that of whose  $A_{\rm M} \le$37,  
the relatively large size 
coronal holes were present  at low-latitudes  and the total solar irradiance
 was very low.
\end{enumerate}

It is believed that the area of a
 sunspot or a
 sunspot group  has a better physical significance   than $R_{\rm Z}$
because the area is a better
 measure (proxy) of  solar magnetic flux than $R_{\rm Z}$~\citep{dg06};
 an area of 130 msh (1 msh
$\approx 3\times 10^6$ km$^2$) corresponds approximately to
 $10^{22}$ Mx (maxwell)~\citep{ws89}.
 The annual sum of the areas of  
sunspot groups   reasonably 
   well 
correlate to the  annual average $R_{\rm Z}$~\citep{jj12b}. 
 The sum of the areas of sunspots and sunspot groups may be representing the 
total  magnetic flux of the sunspots and sunspot groups rather than 
$R_{\rm Z}$.
 It may be worth to note that $R_{\rm Z}$ is not weighted with the areas of
  sunspots and sunspot groups,
  $i.e.$ it gives 
an equal weight to all sunspots and to all sunspot groups regardless of 
their actual size~\citep[$e.g.$][]{cl12}. The yearly mean  $R_{\rm Z}$ 
is determined from the corresponding  monthly means, $i.e.$, the sum of 
 monthly means divided by  number of months 
(it seems to be always 12, in spite  of  in some months the values of the 
sums are zero, $i.e.$ absent of sunspots and sunspot groups). 
Here the  yearly mean size of sunspot groups is directly determined ($cf.$, 
 Eq.. (2)).  
That is, there is a difference in the methods of determinations of
 the yearly mean size of sunspot groups and the yearly mean of $R_{\rm Z}$.  
The annual number of ASGs 
 reasonably well correlate 
to $R_{\rm Z}$ \citep[the   
 variations in NSG, NLG and   NBG  considerably 
differ with that of $R_{\rm Z}$ during 
some solar cycles, see Figs. 2\,--\,4 in][]{jj12a}.  
The sum of the areas of any class
of  sunspot groups
 reasonably correlate to $R_{\rm Z}$.
The good correlations of both the number and the sum of the areas of  
 ASGs with $R_{\rm Z}$  implies that in a given time interval (year) the 
areas of a large 
number of sunspot groups 
 are not drastically different (a few or no outliers). 
It is well-known that the relative frequency of small sunspot 
groups is larger during solar 
minimum than during solar maximum~\citep{mand74}. However, 
 the smallest regions dominate the global flux emergence 
rate~\citep{gs81,zirin87}, and  
small sunspots
 and small sunspot groups  outnumber the large ones~\citep{cl12,jj12a}. 
 This reduces  
  the overall mean sizes of the  sunspots groups  
 in the maximum years (for example, there exits an high anticorrelation 
between the variations that are represented by the solid curves in Figs. 4(a)
 and 6).  
 All these together may be constitute  a possible reason for 
there is  no statistical significant  correlation between the yearly
 mean maximum 
size of sunspot groups in any class and $R_{\rm Z}$ (conclusion~1 above). 
May be for the same reason the annual mean daily rates of growth and decay of 
sunspot groups  do not correlate to the annual  
$R_{\rm Z}$~\citep{jj11}, whereas the annual  rates of 
the growth and decay of sunspot groups  very well correlate to 
 the annual  $R_{\rm_Z}$~\citep{jj12b}.

The existence of approximate 90-year (Gleissberg cycle) and 
160\,--\,270 year~\citep{schove83}  periodicities 
in sunspot activity is known. In Paper~I these periodicities  were seen
 (visualized) 
 in the variations of the numbers (counts) of LSGs/BSGs and SSGs 
(see also Fig.~5), respectively. 
 The existence of  approximate 130-year periodicity (which is seen here in
 the variations of the 
mean maximum size of SSGs, conclusion~3 above) in some solar activity indices
 is also known.
 In addition,  the period of the well-known Gleissberg cycle 
is reported to be very wide, 60\,--\,140 years~\citep{nag97,ogur02}. 
 The aforementioned all  periodicities
 seem to be related  to the
 subharmonics of the 22-year solar
magnetic  cycle~\citep{atto90}.
\cite{gj90,gj92b,gj95} and \cite{gj92a}, from the spherical Fourier analysis of 
a large set of sunspot group data,  found that solar variability  may be the
 resultant of the superpositions of global modes of
 solar magneto-hydrodynamic oscillations. 
\cite{jose65} connected 
 $\approx$ 180-year period of solar magnetic 
activity  to  the 179-year period of the  Sun's motion 
around the solar system barycenter. \cite{jj05} found that 
 violations of the Gnevyshev and Ohl rule or G-O rule~\citep{go48} in sunspot 
cycles may be related  to the Sun's retrograde motion around the solar system 
barycenter, and he has suggested  a role of the Sun's spin-orbit coupling 
 in solar variability.    
 Nevertheless,
the origin of long-term  variations in  solar cycle is still an  unsolved 
problem~\citep[$e.g.$,][]{tan11}. 

 The large/big/strong 
sunspot groups/magnetic-field-regions 
may be associated with a deep dynamo mechanism, whereas the small/weak 
 sunspot groups/magnetic-field-regions 
may be associated with  a near surface dynamo 
mechanism~\citep{gf79,jg97,kmh02,siva03,sg04,meu05,pevt11}. 
 The small sunspot groups may be 
the fragmented or the branched parts of the 
large/big sunspot groups~\citep{jj03}. 
The 130-year and  44-year periodicities 
seen in the variations of the mean maximum sizes of SSGs  and LSGs/BSGs
(conclusion~3 above)  may
 be related to the near surface (local)  and deep (global) dynamo mechanisms,
 respectively.
 
The conclusion~4 above,     is
 consistent with the result that  deficiency of the very small sunspots 
since the activity maximum around 2000~\citep{cl11,cl12},
 and it may be related to a very low growth rate   of
sunspot groups during this period  of cycle~23~\citep{jj11}.
The area of a large/small sunspot group (magnetic flux) decreases in a 
fast/slow rate 
and in fact, it seems the amounts of growth and decay of the magnetic flux
of the sunspot groups  in a given
time interval  depends on (proportional to) the total amount of
the flux in that interval~\citep{jj12b}. The growth of sunspot groups (largely 
contributed by the emergence of new magnetic flux)  may be 
mostly related to the generation mechanism of solar activity, whereas the 
decay of sunspot groups 
 may be related to the   magnetic flux diffusion and 
magnetic reconnection processes.
 The total amount of flux of cycle~23 is relatively small (compared to the 
recent cycles) and it may 
be related to the slow growth of sunspot groups  during this 
cycle~\citep[relatively 
less emergence of new magnetic flux, which is  in addition due to 
 the low efficiency of dynamo, my be due to a strong  effect of the 
Coriolis force on the rising/emerging magnetic flux,][]{fan93}.
The average size of BSGs/LSGs  is relatively small in this 
cycle (see Figs. 4(b) and 4(c)).  \cite{klck11} found that  the average area of 
 a sunspot that belongs to  a large sunspot group decreased 
in solar cycle~23, and 
interpreted this result as the large sunspot groups of cycle~23 
 could be more complex and may  have 
a large number of small sunspots.   
 The  arguments above imply  that  on  the average the sunspot 
groups of cycle~23 are small due to  a low supplement of  
newly emerging magnetic flux, a contradiction to the aforementioned 
conclusion of~\cite{klck11}.   
 The relatively long length  of this cycle may be related  
to the slow growth rate~\citep[and also relatively slow decay/fragmentation 
of BSGs/LSGs  of this cycle because of their relatively small 
size,][]{jj12b}. The relatively slow rate of the fragmentation may be also 
contributed for the less number of small sunspot groups in this 
cycle. It may be interesting to note here   
 that the cycle pair~(22, 23) violated 
the G-O rule mainly due to a large scarcity of SSGs in 
cycle~23~\citep{jj12a}.   
 All these may be
related to the different behavior of the mean merdional motion of the sunspot 
groups, $i.e.$   during this
 cycle the mean 
 meridional motion of sunspot groups was high and its  behaviour 
 was quite different,  south-bound
 during minimum years and north-bound during maximum years~\citep{jj10}. 

\vspace{0.5cm}
%\begin{acknowledgments}

\noindent{\large \bf Acknowledgments}

\vspace{0.3cm}
 {The author is thankful to the two anonymous referees for useful comments 
and suggestions. 
The author  thanks the organizing committee of the 39th COSPAR 
Scientific Assembly, 14\,--\,22 July 2012, Mysore, India,   for providing 
 him a partial financial support to attend and   present, in the 
scientific events  D2.3 and D2.5 of this assembly,  some of  
 the results which are reported here.}  
%\end{acknowledgments}

{}
\clearpage

\begin{figure}
\centerline{\includegraphics[width=\textwidth]{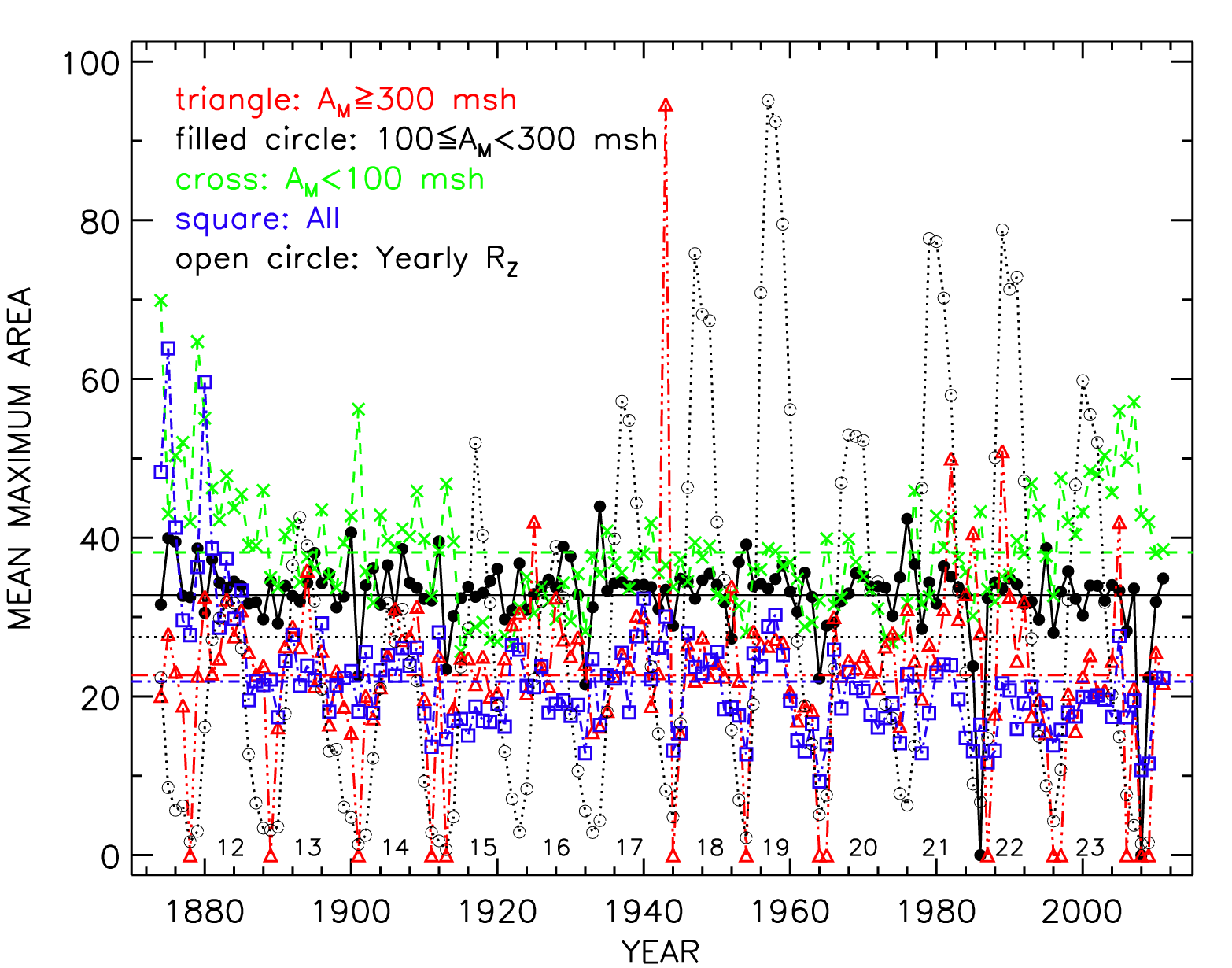}}
\caption{Variations in 
$\bar A_{\rm SSG}$, $\bar A_{\rm LSG}$,  $\bar A_{\rm BSG}$,
 and $\bar A_{\rm ASG}$, $i.e.$   
      the annual mean maximum sizes of the  
small (cross-dashed curve, green), 
large (filled circle-solid curve, black), big (triangle-three dotted-dashed
 curve, red), and all (square-one dotted-dashed curve, blue) spot groups  divided by 1, 5, 20, and 4, 
respectively, during the period 1874\,--\,2011 (Note: a value equal to zero
implies the absent of sunspot groups in the corresponding class, 
  the data in 1874 and
 2011 are incomplete). The open circles connected by the dotted lines
 represents the values   of 
$R_{\rm Z}$ divided by 2. 
 The horizontal lines represent the respective mean values. Near the 
 maximum of each cycle the corresponding Waldmeier cycle number 
is indicated. (For interpretation of the references to color in this figure
 legend, the reader is referred to the web version of this article.)}
\end{figure}

\begin{figure}
\centerline{\includegraphics[width=\textwidth]{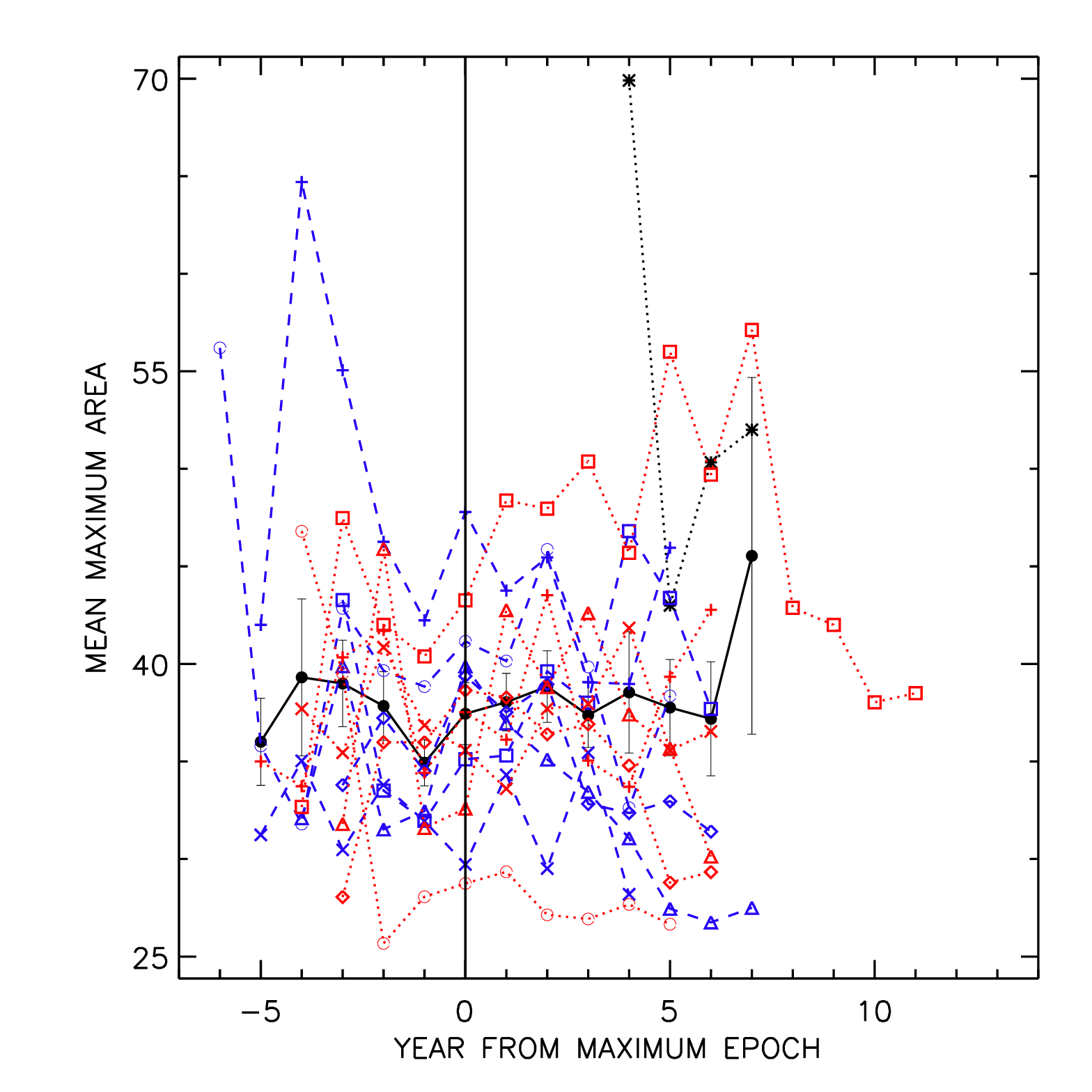}}
\caption{The annual mean maximum sizes  of 
  the small sunspot groups ($\bar A_{\rm SSG}$)
{\it versus} the year from the maximum epoch of the solar cycle.
The blue (dashed curves)  and red (dotted curves) colors are used
 for even- and odd-numbered
cycles, respectively. (For the sake of clarity  black color 
 is used for cycle~11.)  Different symbols are used for
different cycles (numbers are given in the parentheses): asterisks (11),
pluses (12 and 13),
open-circles (14 and 15), crosses (16 and 17), diamonds (18 and 19),
triangles (20 and 21), and squares (22 and 23).
 The filled circles connected by the solid lines represent
 the mean solar cycle variation determined from the  values of annual means.
 The error bars represent  the standard error. 
 There is only one
data point  at  years -6 (beginning of cycle~14), 8 (end of cycle~23) and
9\,--\,11 (first three years of cycle 24).
 (For interpretation of the references to color in this figure legend, the
reader is referred to the web version of this article.)}
\end{figure}

\begin{figure}
\centerline{\includegraphics[width=\textwidth]{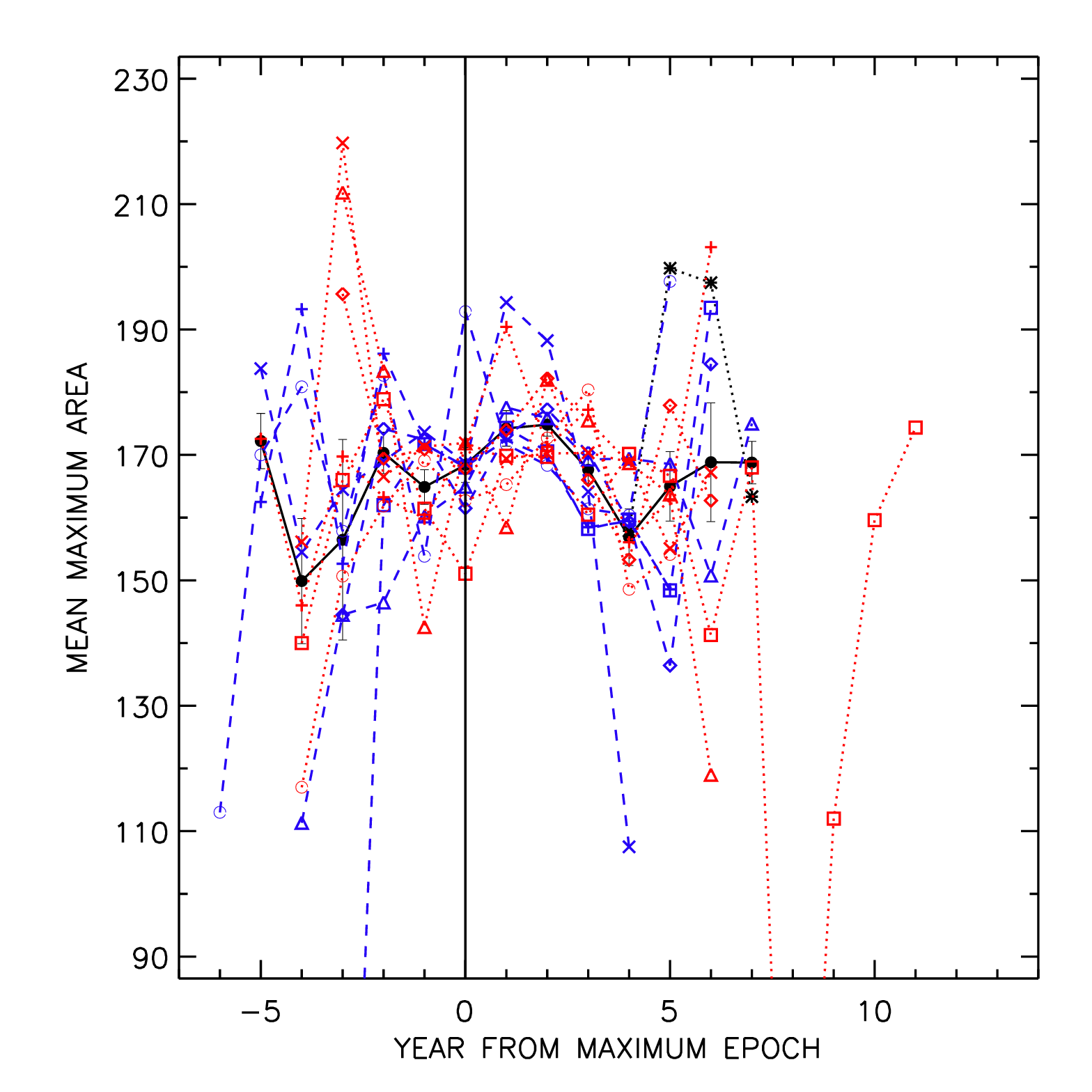}}
\caption{The same as Fig.~2 but for the mean maximum size  of large 
sunspot groups ($\bar A_{\rm LSG}$).  The blue square
 (connected by dashed line)at year  $-3$
 of cycle~22 and the red square (connected by dotted line)
 at year $+8$  of cycle 23
  can't be seen because the corresponding values are  equal to 
zero due to the absent of LSGs.
 (For interpretation of the references to color in this figure legend, the
reader is referred to the web version of this article.)}
\end{figure}

\begin{figure}
\includegraphics[height=6.0cm,width=5.5cm]{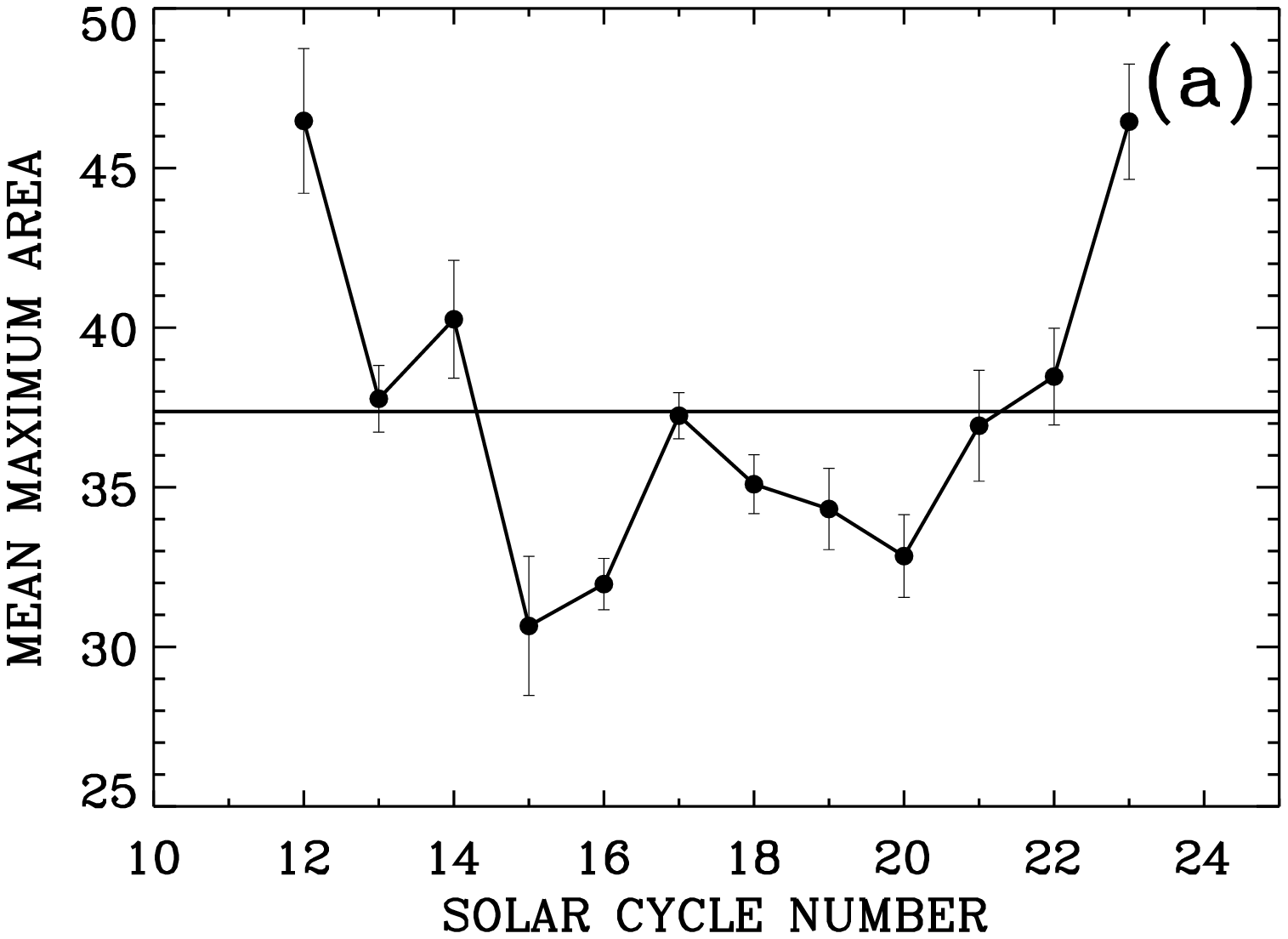}
\includegraphics[height=6.0cm,width=5.5cm]{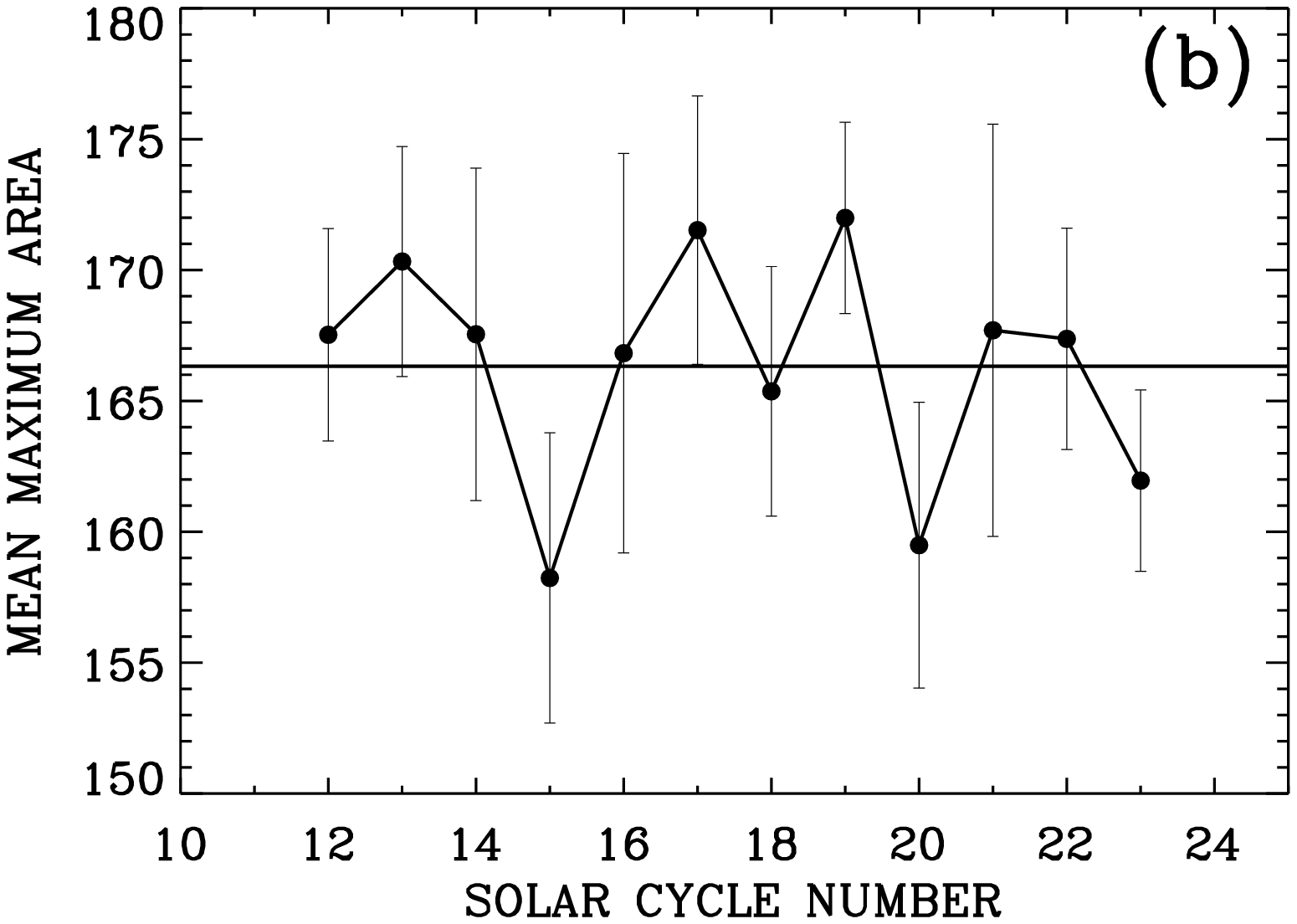}
\includegraphics[height=6.0cm,width=5.5cm]{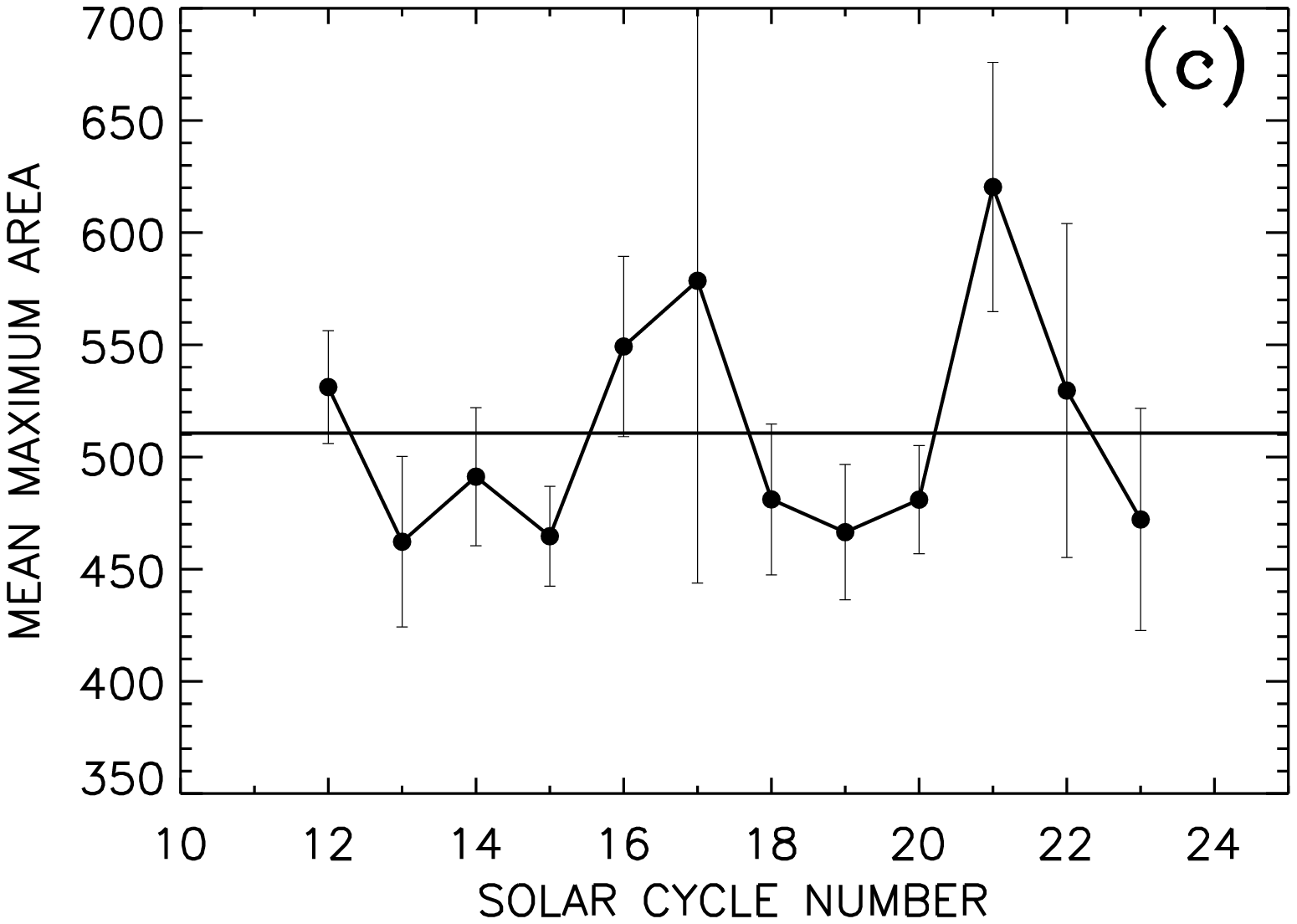}
\includegraphics[height=6.0cm,width=5.5cm]{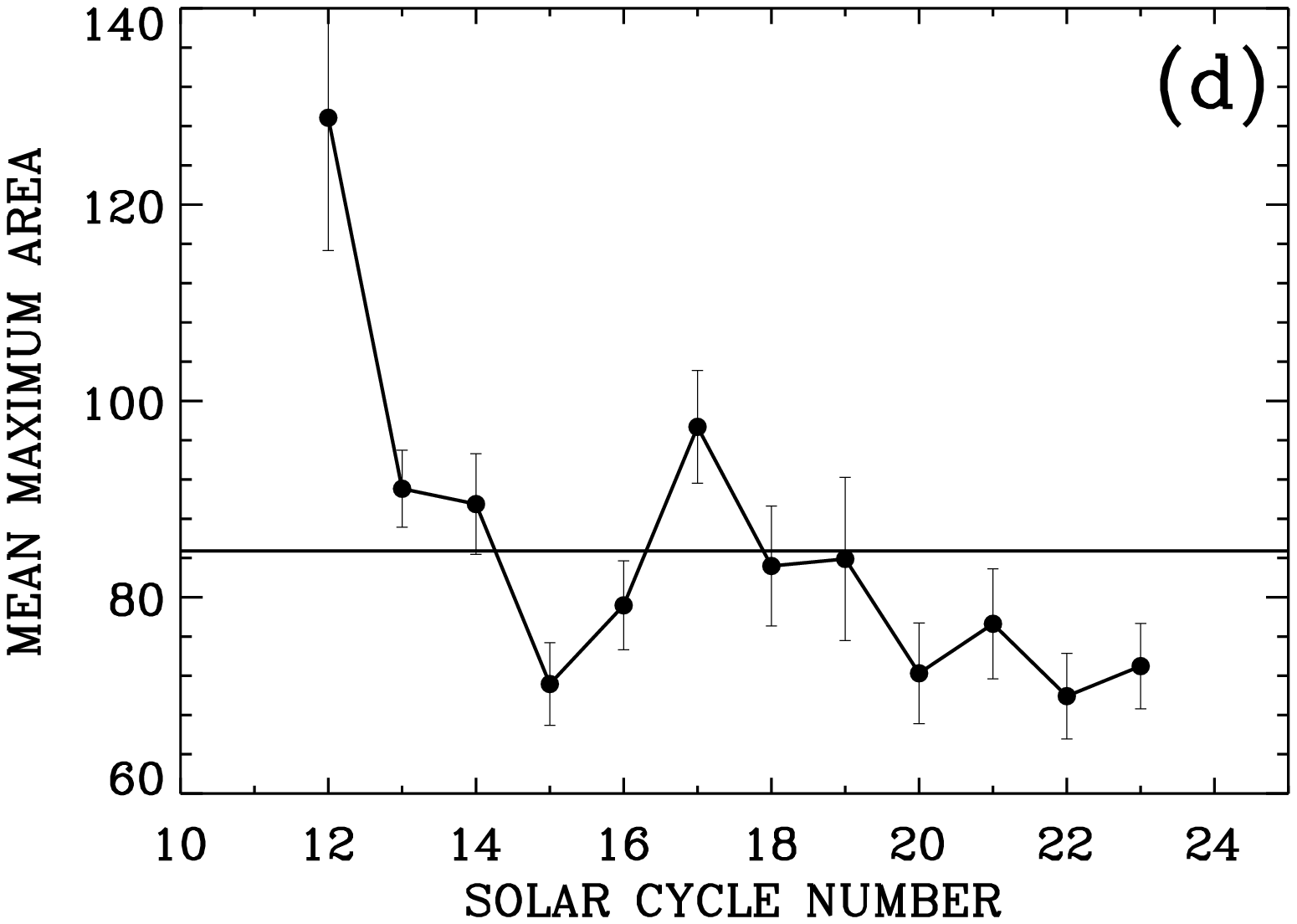}
\caption{Varitions in $\bar A_{\rm WCN}$, $i.e.$ cycle--to--cycle
 variations in the mean maximum sizes (in msh)  of
  (a) 
small sunspot groups (SSGs), 
(b) large sunspot groups (LSGs), (c) big sunspot groups (BSGs), and 
(d) all sunspot groups.
The error bars represent  the standard error.}
\end{figure}

\begin{figure}
\centerline{\includegraphics[width=8cm]{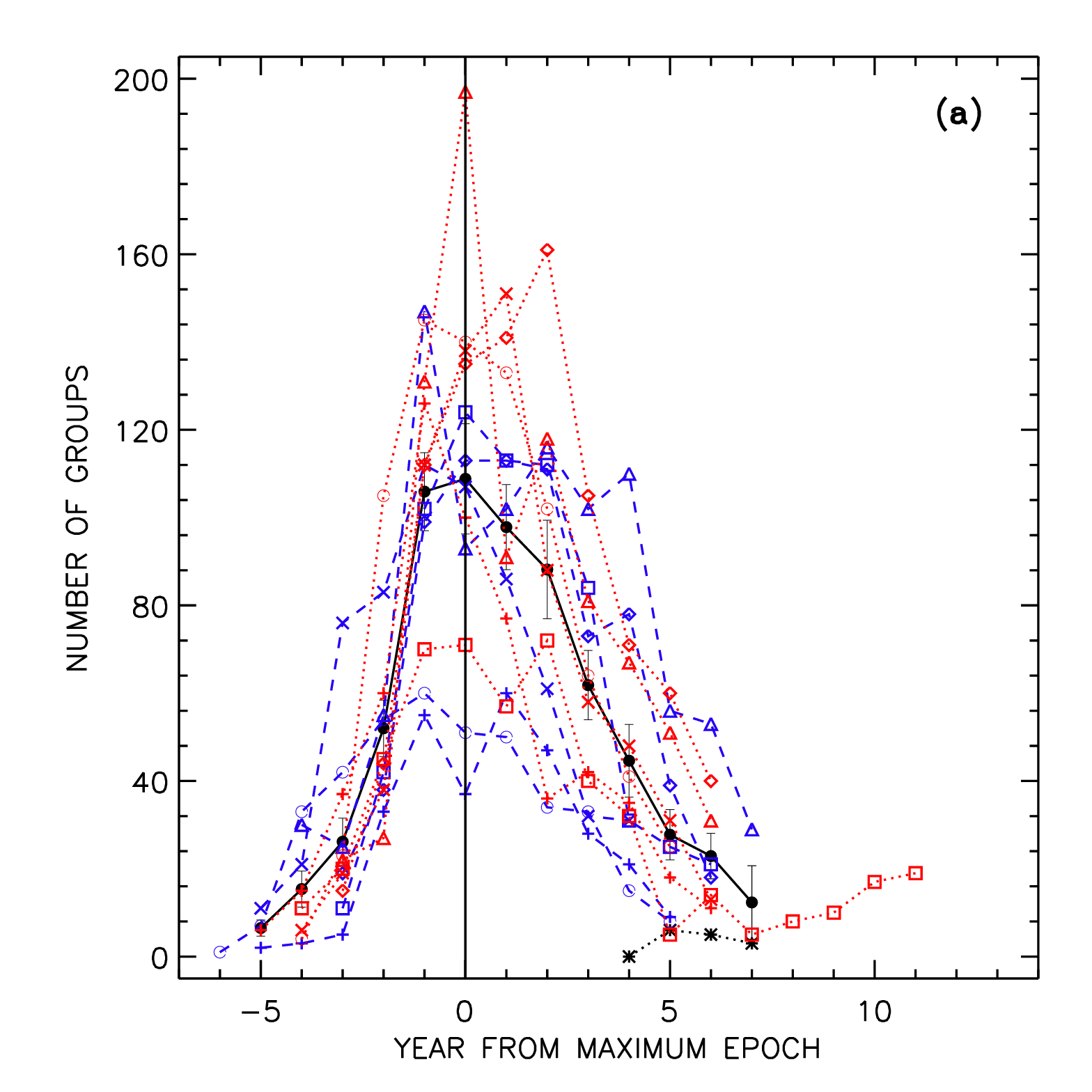}}
\centerline{\includegraphics[width=8cm]{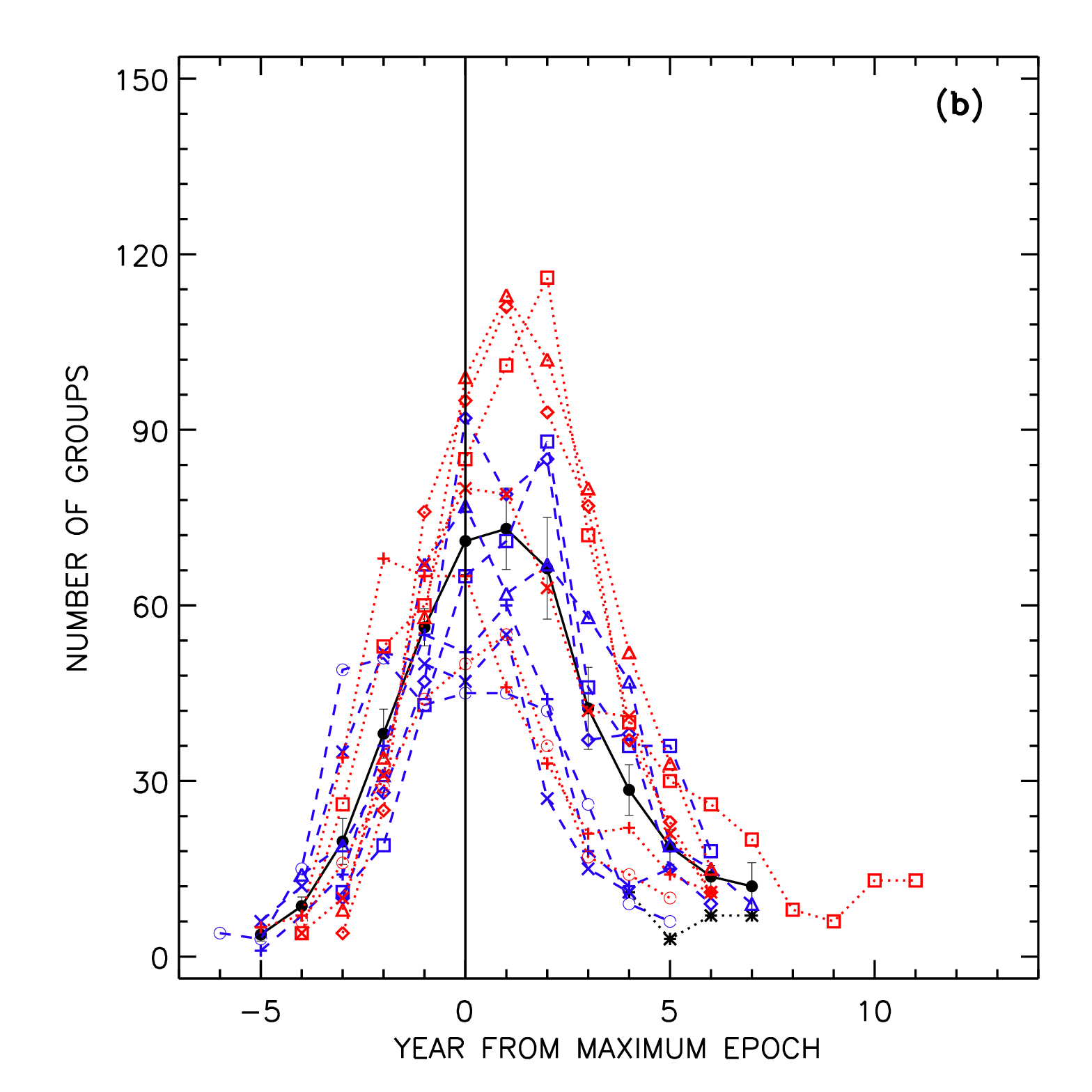}}
\caption{
Solar cycle  variations in the annual numbers of the small sunspot groups
 whose (a)  $A_{\rm M} \le 37$ msh  and 
 (b) $ 37 < A_{\rm M} \le 100$ msh. The different
 symbols and colors represent the different cycles, the same as in Figure~2. 
 (For interpretation of the references to color in this figure legend, the
reader is referred to the web version of this article.)}
\end{figure}

\begin{figure}
\centerline{\includegraphics[width=\textwidth]{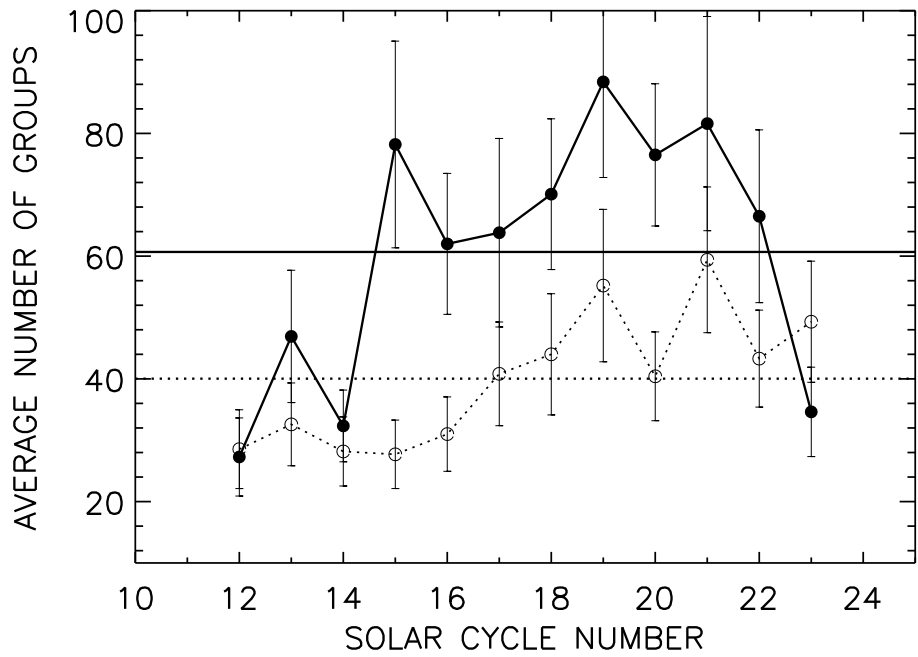}}
\caption{
Cycle-to-cycle variations in the mean annual numbers of small sunspot groups
 whose   $A_{\rm M} \le 37$ msh (filled circle-solid curve) and 
$ 37 < A_{\rm M} \le 100$ msh (open circle-dotted curve.} 
\end{figure}

\begin{figure}
\centerline{\includegraphics[width=\textwidth]{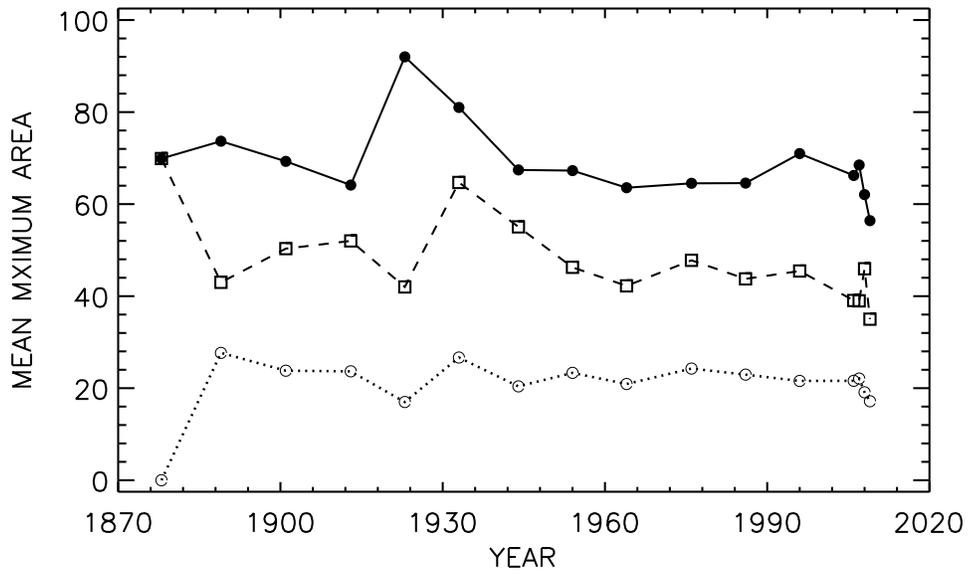}}
\caption{Plots of the annual mean sizes of SSGs 
 whose  $A_{\rm M} \le 37$ msh 
(open circle-dotted curve) and $37 < A_{\rm M} \le 100$ msh 
(filled circle-solid curve), and all SSGs whose $A_{\rm M} \le 100$ msh 
(squares-dashed curve)  
 during the preceding minimum epochs of cycles~12\,--\,23 and around the 
 minimum between cycle~23 and 24 versus the corresponding 
years, 1878, 1889, 1901, 1913, 1923, 1933, 1944, 1954, 1964, 1976, 1986,
1996, 2006, 2007, 2008, and 2009.}
\end{figure}

\end{document}